\documentclass[pra,aps,amsmath, amssymb, twocolumn, superscriptaddress]{revtex4-2}

\usepackage{graphicx}
\usepackage[unicode=true, breaklinks=false, pdfborder={0 0 1}, backref=false, colorlinks=true, linkcolor=blue, citecolor=blue, urlcolor=blue]{hyperref}
\usepackage{amsmath, amssymb, braket}
\usepackage{bm}           
\usepackage{bbm}

\usepackage[authormarkup=none]{changes}
\definechangesauthor[name={Stefan}, color=purple]{n}
\definechangesauthor[name={Stella}, color=orange]{s}

\newcommand{\la}{\langle}
\newcommand{\ra}{\rangle}

\newcommand{\Order}{\mathcal{O}}
\newcommand{\cL}{\mathcal{L}}
\newcommand{\cD}{\mathcal{D}}
\newcommand{\cE}{\mathcal{E}}

\newcommand{\da}{\dagger}

\newcommand{\Op}[1]{\hat{#1}}

\newcommand{\oL}{\Op{L}}
\newcommand{\oM}{\Op{M}}
\newcommand{\oa}{\Op{a}}
\newcommand{\oJ}{\Op{J}}
\newcommand{\oH}{\Op{H}}
\newcommand{\oD}{\Op{D}}
\newcommand{\oA}{\Op{A}}

\newcommand{\oK}{\Op{K}}

\newcommand{\oB}{\Op{B}}
\newcommand{\oU}{\Op{U}}
\newcommand{\oV}{\Op{V}}

\newcommand{\oN}{\Op{N}}

\newcommand{\id}{\ensuremath{\mathbbm 1}}

\newcommand{\diff}{\mathrm{d}}

\DeclareMathOperator{\sinc}{sinc}
\DeclareMathOperator{\tr}{tr}

\begin{document}

\title{
Quantum advantage in charging cavity and spin batteries by repeated interactions
}
\author{Raffaele Salvia}
\email{raffaele.salvia@sns.it}
\affiliation{Scuola Normale Superiore, I-56127 Pisa, Italy}
\affiliation{D\'{e}partement de Physique Appliqu\'{e}e,  Universit\'{e} de Gen\`{e}ve,  1211 Gen\`{e}ve,  Switzerland}

\author{Martí Perarnau-Llobet}
\affiliation{D\'{e}partement de Physique Appliqu\'{e}e,  Universit\'{e} de Gen\`{e}ve,  1211 Gen\`{e}ve,  Switzerland}

\author{Géraldine Haack}
\affiliation{D\'{e}partement de Physique Appliqu\'{e}e,  Universit\'{e} de Gen\`{e}ve,  1211 Gen\`{e}ve,  Switzerland}

\author{Nicolas Brunner}
\affiliation{D\'{e}partement de Physique Appliqu\'{e}e,  Universit\'{e} de Gen\`{e}ve,  1211 Gen\`{e}ve,  Switzerland}

\author{Stefan Nimmrichter}
\email{stefan.nimmrichter@uni-siegen.de} 
\affiliation{Naturwissenschaftlich-Technische Fakult{\"a}t, Universit{\"a}t Siegen, Siegen 57068, Germany}

\begin{abstract}

Recently, an unconditional advantage has been demonstrated for the process of charging of a quantum battery in a collisional model. Motivated by the question of whether such an advantage could be observed experimentally, we consider a model where the battery is modeled by a quantum harmonic oscillator or a large spin, charged via repeated interactions with a stream of non-equilibrium qubit units. For both setups, we show that a quantum protocol can significantly outperform the most general adaptive classical schemes, leading to 90\% and 38\% higher charging power for the cavity and large spin batteries respectively. Towards an experimental realization, we also characterise the robustness of this quantum advantage to imperfections (noise and decoherence)  considering implementations with state-of-the-art micromasers and hybrid superconducting devices. 

\end{abstract}
	
	\maketitle

\section{Introduction}

One central aim of the field of quantum thermodynamics is to improve and fuel thermodynamic processes via quantum resources~\cite{Goold2016,Vinjanampathy2016,Mitchison2019,myers2022quantum}. 
 Promising results  include enhancements in cooling~\cite{Brunner2014,Mitchison2015,Brask2015,Hofer2016,Mitchison2018,Holubec2018,Dorfman2018,Maslennikov2019}, as well as in the power~\cite{Harbola_2012,Uzdin2015,Manzano2016,Klatzow2019,Rodrigues2019,Hammam2021},  efficiency~\cite{Scully2003,Huang2012,Rossnagel2014,Correa2014,Santos2021,Hammam2022}, and reliability~\cite{Agarwalla2018,Guarnieri2019,Liu2019,Ehrlich2021,Menczel2021,Kalaee2021,RignonBret2021} of quantum engines   compared to their classical counterparts. Yet, it is often debated if the potential gains outweigh the cost of preparing the quantum resources, 
 and whether the same output can be simulated by classical means~\cite{Nimmrichter2017,Gonzlez2019,Bumer2018,Lostaglio2020a}.
One approach to overcome these criticisms is to 
devise quantum protocols that overcome \emph{arbitrary} classical strategies given some thermodynamically relevant figure of merit (and a finite set of available resources, like time or energy), in  analogy with quantum advantages developed in quantum metrology~\cite{Giovannetti2011} or quantum computation~\cite{Bharti2022}. 

 A well-known set-up to investigate  
quantum advantages in thermodynamics are quantum batteries, i.e., small quantum systems that store and provide energy~\cite{Campaioli2018,Bhattacharjee2021}. Following pioneering works~\cite{Alicki2013,Hovhannisyan2013,Binder2015}, it has been rigorously proven that entangling operations enable faster battery charging  given a collection of quantum batteries~\cite{Campaioli2017,JuliaFarre2020,Gyhm2022}. In such collective processes, the role of dissipation~\cite{Barra2019,Farina2019,Kamin2020,Hovhannisyan2020,Tabesh2020,Ghosh2021,Zakavati2021}, many-body interactions~\cite{Le2018,Rossini2019,Rossini2020,JuliaFarre2020,Rosa2020,Ghosh2020,Peng2021,Zhao2021,Lu2021,Barra2022,Salvia2022,Ghosh2022}, and energy fluctuations~\cite{Friis2018,McKay2018,PerarnauLlobet2019,Pintos2020,Crescente2020,Caravelli2020random,Salvia2021} has also been investigated.   In parallel, various types of quantum systems have been considered as quantum batteries, ranging from qubit ensembles~\cite{Binder2015,Ferraro2018,Andolina2019,Andolina2019II,Gherardini2020,Santos2020,Quach2020,Caravelli2020,Landi2021} and ladder models~\cite{Santos2019,Pintos2020,mitchison2020charging,Lobejko2020,Seah2021,LipkaBartosik2021,obejko2021} to oscillators and flywheels~\cite{Levy2016,Seah2018,Andolina2018,Tabesh2020,Tirone2021}. 
One of the prime candidates for experimental proofs-of-principle is the Dicke-model battery~\cite{Fusco2016,Ferraro2018,Andolina2019,Andolina2019II,Pirmoradian2019,Dou2022}, in which an ensemble of qubits is charged by a collectively coupled oscillator mode (see also related proposals in waveguide QED set-ups~\cite{Monsel2020,Maffei2021}). Recent efforts in cavity- and circuit-QED have brought about the first results \cite{Hu2021,Quach2022}. 

\begin{figure}
	\centering
	\includegraphics[width=\columnwidth]{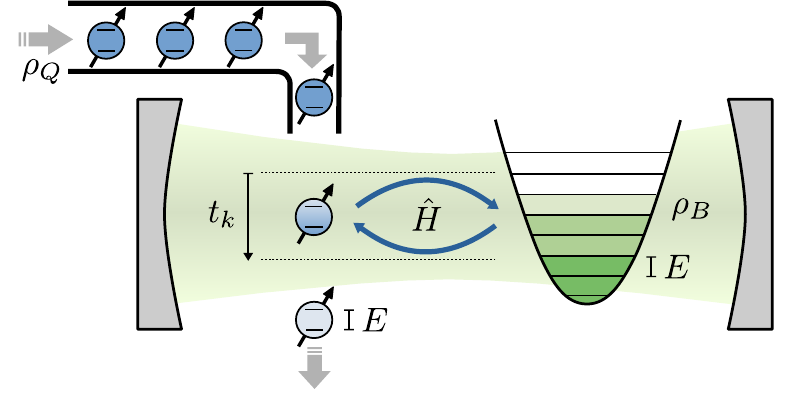}
	\caption{Sketch of a micromaser as a quantum battery: Collisional charging of a harmonic oscillator battery (e.g.~a cavity mode) by a sequence of identically prepared qubits. Each qubit carrying energy charge in its quantum state $\rho_Q$ is allowed to exchange resonant excitations of energy $E$ with the battery state $\rho_B$ via the Hamiltonian $\oH$. In adaptive protocols, the interaction times $t_k$ can vary with the qubit number~$k$.}
	\label{fig:frameworkI}
\end{figure}

In a recent article, some of us investigated quantum advantages in battery charging which are not of collective nature~\cite{Seah2021}. We considered a collisional model where a stream of non-equilibrium qubit units sequentially interact with an  energy ladder, which stores the energy thus playing the role of the battery.  
In this  model,  we showed that quantum-coherent processes (where the qubits contain quantum coherence in the energy basis) can outperform arbitrary classical incoherent processes in terms of charging power. However,  the interaction between the qubit units and the energy ladder considered in~\cite{Seah2021} was  idealised,  being independent  of the particular energy state of the ladder (see details below).  In this work, we extend these promising results to  realistic physical set-ups. 
For that, we replace the ideal energy ladder of~\cite{Seah2021} by two different physical models of quantum batteries: (i)~a quantum harmonic oscillator, and (ii)~a large spin system. These two models can be realised  
with state-of-the-art  experimental platforms, including  (i) a micromaser  (i.e. a cavity QED  charged by an atomic ensemble~\cite{Sayrin2011}) and a circuit QED set-up~\cite{Linpeng2022}, and for (ii)  hybrid circuit QED-magnetic systems~\cite{Zhang2015,Wolski2020}. 
For both models, we show that a suitably chosen coherent charging protocol can outperform the best incoherent adaptive charging protocol in terms of power, even in the presence of environmental damping or decoherence for state-of-the-art experimental values. 

The manuscript is organized as follows. In Sec.~\ref{section:framework}, we describe the collisional model and relevant quantities.  In  Sec.~\ref{section:jaynes-cummings},
we  analyse the case of a harmonic oscillator battery. We derive an upper bound on the charging power for any adaptive incoherent protocol and demonstrate how to beat it with coherent protocols. We discuss possible realizations in cavity-QED and circuit-QED setups and show that the coherent quantum advantage can withstand weak damping due to photon loss. In Sec.~\ref{section:spin}, we present similar results for a large spin battery, which could be realized with hybrid circuits involving a superconducting qubit that would sequentially interact with a magnon~\cite{Wolski2020}. Finally, we conclude in Sec.\ref{section:conclusions}. 

\section{Framework}
\label{section:framework}

We  first present some generic considerations for the sequential charging of the quantum battery by identical qubits, which serves as the basis for our treatment of oscillator and spin batteries. 
Let us consider the battery Hamiltonian~$\oH_B = E\sum_n n|n\ra\la n| = E \oN$ describing a uniform, bounded or unbounded, energy ladder in steps of~$E>0$. Starting in the ground state of empty charge, $\rho_B (0)=|0 \ra \la 0| $, the battery shall be charged by a sequence of resonant qubits with $\oH_Q = E |e\ra\la e|$, as sketched in Fig.~\ref{fig:frameworkI}. 
Each qubit is prepared in a state
    \begin{align}
          \rho_Q =& q |g \ra \la g| + (1-q) |e \ra \la e|  \nonumber\\ &+c \sqrt{q(1-q)} \left(  e^{-i\alpha} |g \ra \la e| + e^{i\alpha} |e \ra \la g| \right), \label{eq:qubitState}
    \end{align}
where $q\in [0,1]$ is the ground-state population, $c\in [0,1]$ is  the amount of coherence in the state, and $\alpha \in [0,2\pi]$ is a phase. Along this work, we make a distinction between quantum coherent protocols (where $c\neq 0$) and incoherent or classical protocols ($c=0$). 

In the $k$th step of the battery charging process, a qubit unit and the battery interact for a time $t_k$ via the resonant exchange Hamiltonian 
   \begin{equation}\label{eq:Hint}
    \oH = \Omega \left(\oA \otimes |e\ra\la g| + \oA^\da \otimes |g\ra\la e|\right),
\end{equation}
at a fixed Rabi frequency $\Omega$. 
Here, $\oA$ represents a model-dependent ladder operator satisfying $\la m|\oA|n\ra \propto \delta_{m,n-1}$. In our previous work Ref.~\cite{Seah2021}, we considered an ideal energy ladder with $\oA= \sum_{k=1} \ket{k-1} \bra{k}$, whereas here we will focus on the following two physically relevant scenarios: 
\begin{itemize}
    \item Quantum harmonic oscillator, with  $\oA$ a bosonic annihilation operator, 
    $\oA=\oa=\sum_{k=1} \sqrt{k} \ket{k-1} \bra{k}$.
     \item Large spin with $\oA=\oJ_-$, the exact definition being introduced in  section~\ref{section:spin}.
\end{itemize}
Note that the total energy is preserved during the interaction process, $[\oH,\oH_B+\oH_Q]=0$,  so that no additional work must be consumed in this process. 

In the interaction picture with respect to $\oH_B+\oH_Q$, the state $\rho_B$ of the battery evolves at the $k$th charging step as: 
    \begin{align}
     \rho_B (k+1) = \tr_Q \left\{ e^{-i t_k \oH} \rho_B (k) \otimes \rho_Q e^{i t_k \oH} \right\},
     \label{eq:evolutionrhoB}
    \end{align}
Such interaction processes are applied for a total available time $\tau$, i.e.,
\begin{align}
\tau= \sum_{k=1}^K t_k,
\label{eq:tau}
\end{align}
where $K$ is the total number of qubits. We regard the charging time $\tau$ as the limiting resource, whereas~$K$ and~$\{ t_1,...,t_K \}$ can be freely chosen with the only constraint of Eq. \eqref{eq:tau}. In what follows, it will be convenient to introduce  the dimensionless interaction times (swap angles): 
\begin{align}
    \theta_k=\Omega t_k. 
\end{align}

Our goal is to maximise the final average energy of the quantum battery, $\bar{E}(K)=\tr \{ \rho_B(K) \oH_B \}$,  under the time constraint~$\tau$. In other words, we wish to maximise the charging \textit{power}
\begin{align}
    \bar{\mathcal{P}}= \frac{\bar{E}(K)}{\tau}. 
    \label{eq:DefPower}
\end{align}
We will consider this optimization for both classical ($c=0$) and quantum ($c\neq 0$) protocols. For convenience, we also introduce the transient charging power of the $k$-th step,
\begin{align}
    \mathcal{P}(k)=\frac{\bar{E}(k)-\bar{E}(k-1)}{t_k}. 
\end{align}

Although \eqref{eq:DefPower} is the central figure of merit of this work, we note that not all energy in $\rho_B(K)$ can be extracted as useful work. Some fraction of $\bar{E}(K)$ will be \textit{passive} energy in the form of heat, since the battery state has in general non-zero entropy. This  motivates us to also compute the ergotropy $\mathcal{E}(K)$, which quantifies the amount of work that can be extracted from $\rho_B (K)$ in a cyclic Hamiltonian process~\cite{Allah2004,Goold2016}. Given a quantum state $\rho$ with Hamiltonian $\oH$, the ergotropy $\mathcal{E}$ is defined by $\mathcal{E}= \tr{\rho \oH} - \tr{\rho^{\rm passive} \oH}$, where $\rho^{\rm{passive}}= \sum_n r_n |n\ra\la n|$ with $[\rho^{\rm passive},\oH]=0$ and $\{r_n\}$ the eigenvalues of the initial state in descending order. Hence, the \textit{useful} battery charge is always upper-bounded by the average energy, $\mathcal{E}_B \leq \bar{E}_B$, and equality only holds for pure battery states in our case. 

Before moving forward, we also note that the evolution of the battery \eqref{eq:evolutionrhoB} can be effectively described by a master equation model. While this approach is not strictly required to derive the main results presented in this work, it is both physically insightful and technically useful (e.g., for assessing the quantum-coherent evolution of the battery). We present it in detail in  Appendix \ref{app:dissipatori} for the interested reader.

\section{Harmonic oscillator battery}
\label{section:jaynes-cummings}

\begin{figure}
	\centering
	\includegraphics[width=\columnwidth]{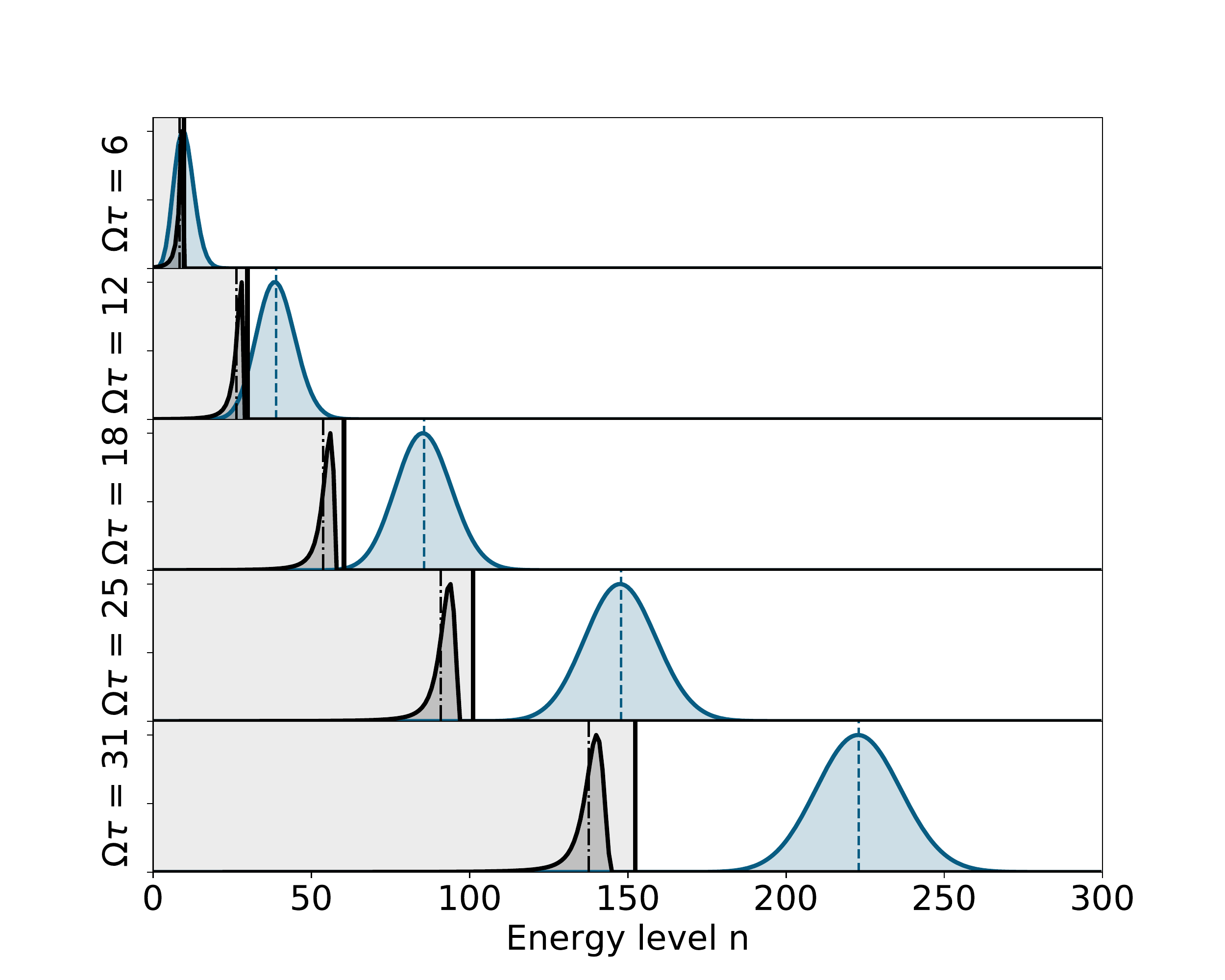}
	\caption{Energy distribution of an oscillator battery at various charging times $\tau$, comparing an exemplary coherent charging protocol (blue) to an adaptive incoherent protocol (black). The vertical lines mark the average energies. The former assumes qubits in a uniform superposition ($q=1/2$) and a constant interaction time $\theta=0.01\pi$, whereas the latter operates with fully excited qubits ($q=0$) and optimizes the charging time for maximum transient power in each step. The incoherent protocol beats deterministic charging via full swaps (dashed vertical line), but does not reach the incoherent power bound (grey shade boundary). 
	}
	\label{fig:coh1_distr}
\end{figure}

We first consider the battery to be a quantum harmonic oscillator with Hamiltonian $\oH_B = E \oa^\da\oa $ and interacting Hamiltonian   
    \begin{align}
         \label{eq:Hintho}
    \oH = \Omega (\oa \otimes |e\ra\la g| + \oa^\da \otimes |g\ra\la e|). 
    \end{align}
For this micromaser model, Figure~\ref{fig:coh1_distr} displays snapshots of the collisional battery charging. Indeed, it displays the battery energy distribution at increasing times during exemplary charging protocols.
We compare coherent charging (blue) by weakly interacting qubits in the $|+\ra$-state ($\theta = 0.01\pi$, $q=1/2$) to an adaptive protocol (black) with fully excited qubits ($q=0$), in which we optimize the interaction time $\theta_k$ in each step for maximum transient power. This latter protocol outperforms a deterministic protocol of full excitation swap from each qubit (dashed vertical line), defined by $q=0$ and $\theta_k = \pi/2$ $\forall k$. However, it cannot reach the coherent performance as it must obey an incoherent bound on the charging power (grey shade boundary) that we will work out in the following. Finally,  we also observe that the quantum random-like behaviour described in Ref.~\cite{Seah2021} (see Fig.~2 of~\cite{Seah2021}) is lost due to the different nature of the interaction. 

\subsection{Bound on the incoherent charging power}

We first focus on incoherent charging strategies where $c=0$ in~\eqref{eq:qubitState}. Let us expand~$\oH_B = E \sum_n n |n\ra\la n| $ and recall that initially the battery starts in the ground state,~$\rho_B (0)=|0\ra \la 0|$.  The state~$\rho_B (k)$ remains energy-diagonal under the evolution \eqref{eq:evolutionrhoB}. The evolution of the populations $p(n,k)=\bra{n}\rho_B (k) \ket{n}$  can  be described by the rate equation
\begin{eqnarray}
    p(n,k+1) &=& P_0(n,k)p(n,k) + P_- (n+1,k)p(n+1,k) \nonumber \\
    && + \,\, P_+ (n-1,k) p(n-1,k). \label{eq:MEinc_HO}
\end{eqnarray}
where  $P_- (n,k) = q\sin^2 (\sqrt{n}\theta_k)$, $P_+ (n,k) = (1-q)\sin^2 (\sqrt{n+1}\theta_k)$, and $P_0 (n,k) = 1-P_- (n,k)-P_+ (n,k)$. 

Consider the mean energy gain of the battery at the $k$-th incoherent charge step,
\begin{eqnarray}
\Delta \bar E (k) &=& \bar{E}(k)-\bar{E}(k-1)  \nonumber \\
&=& E\sum_{n=0}^\infty \left[ P_+ (n,k)-P_- (n,k) \right] p(n,k-1) \nonumber \\
&\leq& E\sum_{n=0}^\infty \sin^2 (\sqrt{n+1}\theta_k) p(n,k-1). \label{eq:DeltaE_inc_HO}
\end{eqnarray}
Naturally, this gain is optimal if we consider fully excited charge units ($q=0$), as stated in the last line. Note that we assume that the swap angle $\theta_k$ can be adapted over the course of the sequence by changing the coupling time.

We now consider the transient charging power, 
\begin{eqnarray}
	\label{eq:jensen_maxPowerII}
	\frac{\mathcal{P}(k)}{E\Omega} &=& \frac{\Delta \bar E (k)}{E\theta_k}  \leq 
	 \sum_{n=0}^{\infty} \frac{\sin^2 (\sqrt{n+1}\theta_k)}{\theta_k} p(n,k-1).
\end{eqnarray}
It can be upper bounded by optimising the level-dependent excitation rates over the swap angle (see also~\cite{Andolina2018}),
\begin{eqnarray}
	\max_\theta \frac{\sin^2(\sqrt{n+1}\theta)}{\theta} = R_0 \sqrt{n+1} \; ,
	\label{eq:maxExcitationRate_HO}
\end{eqnarray}
with $R_0 \approx 0.725$ at $\theta \approx 0.742\, \pi/(2\sqrt{n+1})$. 
Inserting this result into \eqref{eq:jensen_maxPowerII} and making use of Jensen's inequality, we arrive at a non-tight bound,
\begin{eqnarray}
	\label{eq:jensen_maxPower}
	\frac{\mathcal{P}(k)}{E\Omega} & \leq& R_0 \sum_{n=0}^{\infty} \sqrt{n+1} p(n,k-1) \nonumber\\
	&\leq& R_0 \sqrt{\sum_{n=0}^{\infty} (n+1) p(n,k-1) } 
	\nonumber\\
	&=& R_0 \sqrt{\frac{\bar{E} (k-1)}{E} + 1}.
\end{eqnarray}
Let us now define a continuous function $\bar E (\Omega \tau) $ of time that interpolates linearly between the values of the mean energy at successive charge steps. Following the derivation of \eqref{eq:jensen_maxPower}, this function must satisfy \begin{equation}
	\frac{\diff \bar E (\Omega \tau) }{\diff \tau} \leq \Omega E R_0 \sqrt{ \frac{\bar E(\Omega \tau)}{E} + 1} \; ,
	\label{eq:meanE_interpolatedODE_HO}
\end{equation}
which can be integrated by separation of variables. Given the initial condition $\bar E(0) = 0$, we arrive at the inequality
\begin{eqnarray}
	\cE (\Omega \tau) \leq \bar E (\Omega \tau) \leq E R_0 \Omega \tau \left( 1 + \frac{R_0 \Omega \tau}{4} \right).
	\label{eq:Ebound_HO}
\end{eqnarray}		
It gives an upper bound for the gain in mean energy (and thus ergotropy) after the charging time $\tau$. Accordingly, it bounds the average charging power \eqref{eq:DefPower} as 
\begin{align}
    \bar{\mathcal{P}} \leq E \Omega R_0 \left( 1 + \frac{R_0 \Omega \tau}{4} \right).
    \label{eq:UppBoundPower}
\end{align}
This bound is shown in Figs.~\ref{fig:coh1_distr},~\ref{fig:coh1} and~\ref{fig:coh1_loss}, where it defines the grey region  potentially accessible by classical charging protocols. 
We discuss the saturability of this upper bound by classical protocols in the next section.


\subsection{Classical strategies}

A natural candidate for an optimal classical strategy is a full-swap protocol: one decreases the interaction time in each charge step $k$ according to $\vartheta_k = \pi/2\sqrt{k}$, such that the battery remains in a pure charge state, $p(n,k) = \delta_{nk}$, with the mean energy and ergotropy climbing up in steps of $E$. 
The corresponding transient power at the $k$-th step is given by $\mathcal{P} (k) = \Omega \, \Delta \bar{E}(k)/\vartheta_k = 2\sqrt{k}\Omega E/\pi$. 
Assuming further that there is no waiting time between subsequent charge steps, the cumulative charging time is
\begin{equation}
	\Omega \tau = \sum_{j=1}^K \vartheta_j = \frac{\pi}{2} \sum_{j=1}^K \frac{1}{\sqrt{j}} = \pi \sqrt{K} - \zeta_{1/2} + \Order\left\{\frac{1}{\sqrt{K}}\right\},
	\label{eq:cumTime_fullSwap_HO}
\end{equation}
with the Riemann zeta value $\zeta_{1/2} \approx 1.46$. Hence the cumulatively averaged charging power becomes $\bar{\mathcal{P}}=\bar{E}(K)/\tau = KE/\tau \approx \mathcal{P}(K)/2 $ for $K \gg 1$. 
Using \eqref{eq:cumTime_fullSwap_HO},  this can be expressed in terms of the total time as:
\begin{align}
    \bar{\mathcal{P}} \approx \frac{E \Omega^2  \tau}{\pi^2}.
\end{align}
For large $\tau$, it reaches $0.77$ of the upper bound \eqref{eq:UppBoundPower} (noting that both expressions grow linearly with $\tau$). 

By numerically optimising over the interaction time, it is possible to find strategies that weakly outperform the full swap in terms of power. The example shown in Figs.~\ref{fig:coh1_distr} and~\ref{fig:coh1}  is the \emph{greedy} incoherent protocol, i.e., the one that maximizes at each step the cumulative averaged power. We conclude that while the bound \eqref{eq:UppBoundPower} is not tight, classical strategies can perform close to it. In what follows, we show that quantum strategies can outperform the classical bound.

\subsection{Coherent charging}
\label{sec:coherent}

\begin{figure}
	\centering
	\includegraphics[width=\columnwidth]{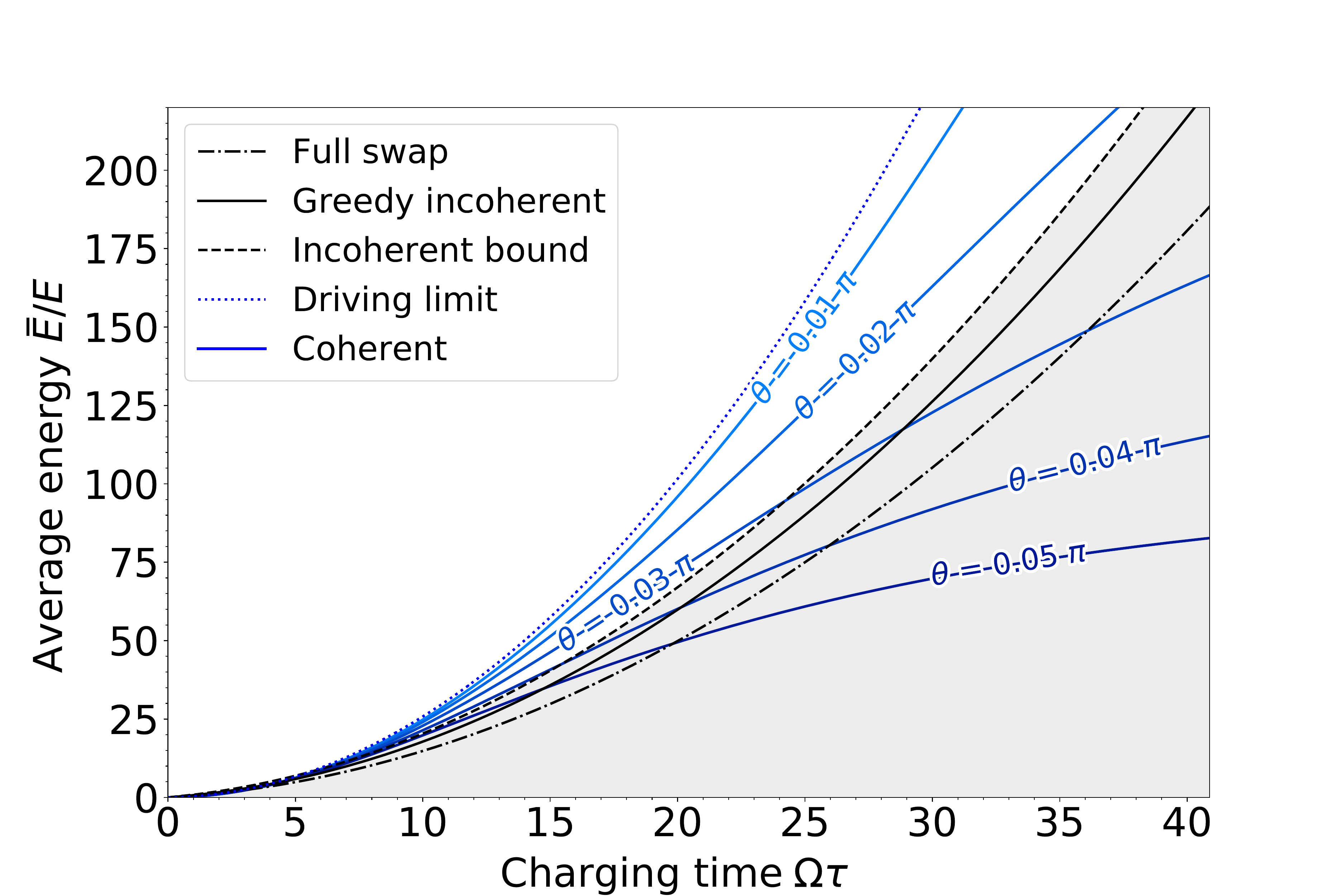}
	\caption{Collisional charging of a truncated harmonic oscillator battery with $250$ levels, using coherent qubits in the $|+\ra$-state ($q=0.5$). Blue solid lines of darker shades show the mean battery energy $\bar{E}$ as a function of charging time $\Omega \tau = K\theta$ for increasing swap angles $\theta$, while the blue dotted line represents the driving limit $\theta \to 0$. For comparison, the black dashed line marks the upper bound for adaptive incoherent charging strategies, closely approximated by a greedy protocol (black solid line) that optimizes the cumulatively averaged power at every charge step using excited qubits ($q=0$). The dash-dotted line shows the deterministic full-swap protocol ($q=0$, $\vartheta_k = \pi/2\sqrt{k}$) in which the battery remains in a pure charge state at all times. }
	\label{fig:coh1}
\end{figure}

One can beat the incoherent bound \eqref{eq:Ebound_HO} in a non-adaptive protocol using pure coherent charge qubits and small swap angles $\theta$. Specifically, for transient charging of an empty battery up to the $N$-th level, a coherent advantage can be achieved in the short-time limit $\sqrt{N+1}\theta \ll 1$ where the charging process approximates a coherent driving term (see details in App.~\ref{app:dissipatori}). The best performance is obtained for equal superposition states with $q=1/2$ and $c=1$ in \eqref{eq:qubitState}, which yield the strongest driving amplitude. 

In Fig.~\ref{fig:coh1}, we show the charging of an oscillator battery with such qubit states (at $\alpha=0$) as a function of charging time $\Omega \tau = K\theta$, for various $\theta$-values and up to at most $N=249$. The incoherent bound (dashed line) and the full-swap protocol (dash-dotted) are clearly beaten for $\theta \lesssim 1/\sqrt{250} \approx 0.02 \pi$, whereas larger swap angles perform worse eventually. 

Ideal performance is reached asymptotically in the short-time limit $\theta \to 0$ (and $K\to \infty$), in which the battery state remains a pure displaced vacuum state at all times, 
\begin{equation}
    \rho_B(K) \xrightarrow[K\to\infty]{\theta\to 0} \oD\left(\frac{\Omega \tau}{2}\right) |0\ra \la 0| \oD\left(-\frac{\Omega \tau}{2}\right),
\end{equation}
with $\oD (\xi)$ the  displacement operator. This follows from the small-$\theta$ expansion \eqref{eq:MEcoh_small} of the coherent charging generator, which we  perform in App.~\ref{app:dissipatori} for generic ladder batteries.  
The corresponding battery charge (dotted line) grows quadratically with time, $\cE (\Omega \tau) = \bar E (\Omega \tau) = E\,(\Omega \tau)^2/4$, resulting in the average charging power
\begin{align}
    \bar{\mathcal{P}} = \frac{E\,\Omega^2 \tau}{4}.
\end{align}
It outperforms the upper classical bound \eqref{eq:UppBoundPower} by a factor $R_0^{-2} \approx 1.90$--- corresponding to a $90\%$ quantum advantage. 
We stress that this is a theoretical limit in which each qubit unit transfers only infinitesimal amounts of energy (and thus wastes most of its useful energy) and the state of the battery remains pure. Nevertheless, in the next section we show that the quantum advantage prevails in state-of-the-art experimental platforms with decoherence and a finite number of qubit units.


\subsection{Photon loss and suitable experimental platforms} \label{sec:experiment}

Several experimental platforms can correspond to a quantum battery modeled as a quantum harmonic oscillator, for instance cold atoms and circuit QED setups. With these experimental platforms, it is important to assess the negative impact of photon loss on the coherent charging process, specifically, the critical loss rate at which the quantum advantage disappears. 
To capture the effect of photon loss, we consider that 
subsequent charging steps are separated by a finite waiting time $t_0$ during which a fraction of the battery charge is lost. Employing the standard bosonic damping channel with rate $\kappa$, the battery state transforms in between subsequent charge steps according to the master equation 
   $ \partial_t \rho = \kappa  \mathcal{D}[\hat{a}] \rho$,
where  $\cD[\oM]\rho = \oM\rho\oM^\da - \{\oM^\da\oM,\rho\}/2$.
This results in the CPTP map \cite{Ivan2011}
\begin{eqnarray}
    \Phi_0 \rho_B &=& e^{\kappa t_0 \cD[\oa]} \rho_B = \sum_{n=0}^\infty \frac{(e^{\kappa t_0}-1)^n}{n!} \oK_n \rho_B \oK_n^\da, \nonumber \\
    \oK_n &=& \oa^n \exp \left(-\frac{\kappa t_0}{2} \oa^\da\oa \right). \label{eq:damping}
\end{eqnarray}
In the following, we employ the commonly used damping parameter $\gamma = 1-e^{-\kappa t_0}$, which approximates $\gamma \approx \kappa t_0$ when small.

\begin{figure}
	\centering
	\includegraphics[width=\columnwidth]{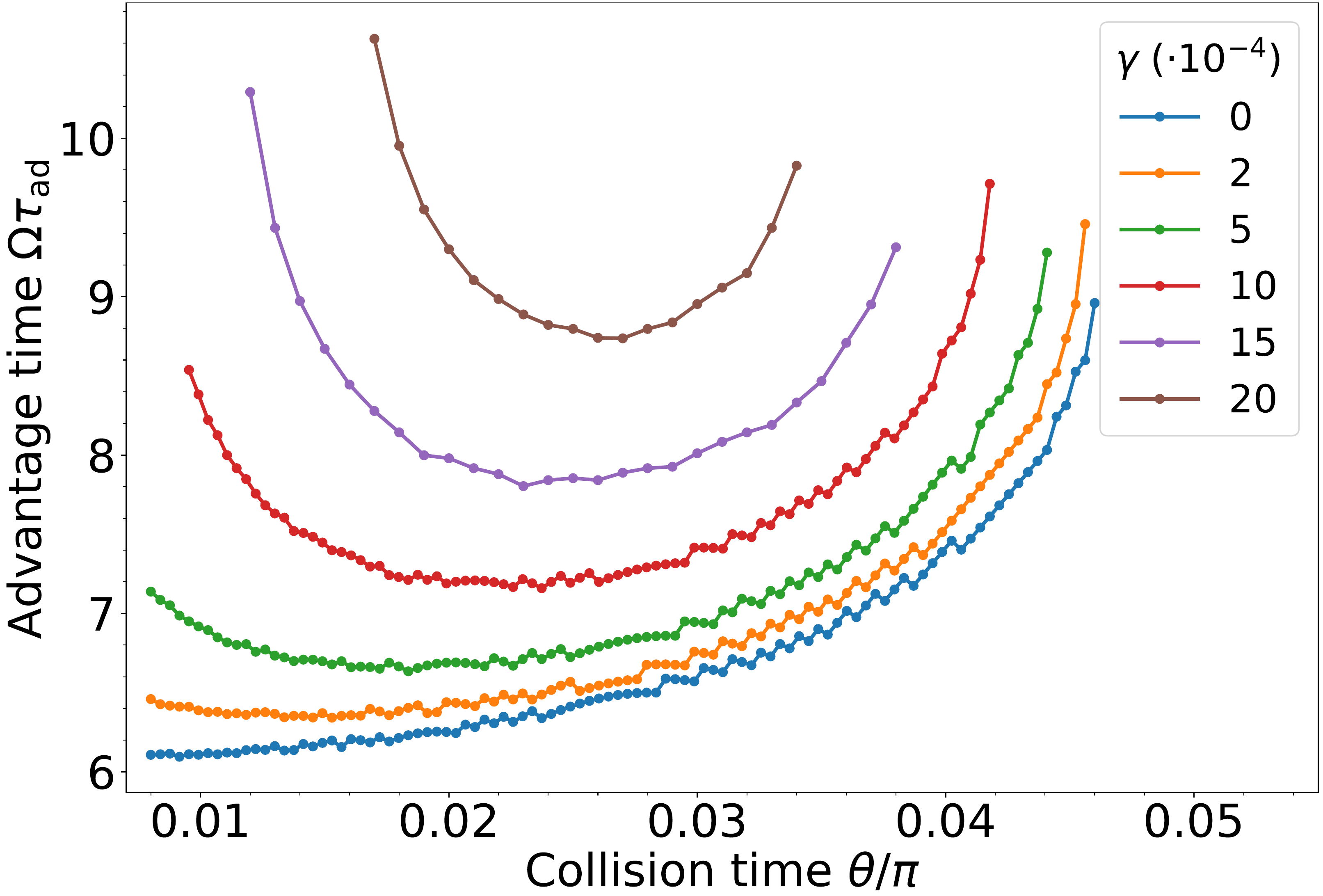}
	\caption{Time $\tau_{\rm {ad}}$ at which the energy of a cavity charged through collisions with constant interaction time $\theta$, and coherent qubits with $q=0.5$, firstly overcomes the incoherent bound \eqref{eq:Ebound_HO}  for different values of the inter-collisional loss $\gamma$.}
	\label{fig:suptim}
\end{figure}

\begin{figure}
	\centering
	\includegraphics[width=\columnwidth]{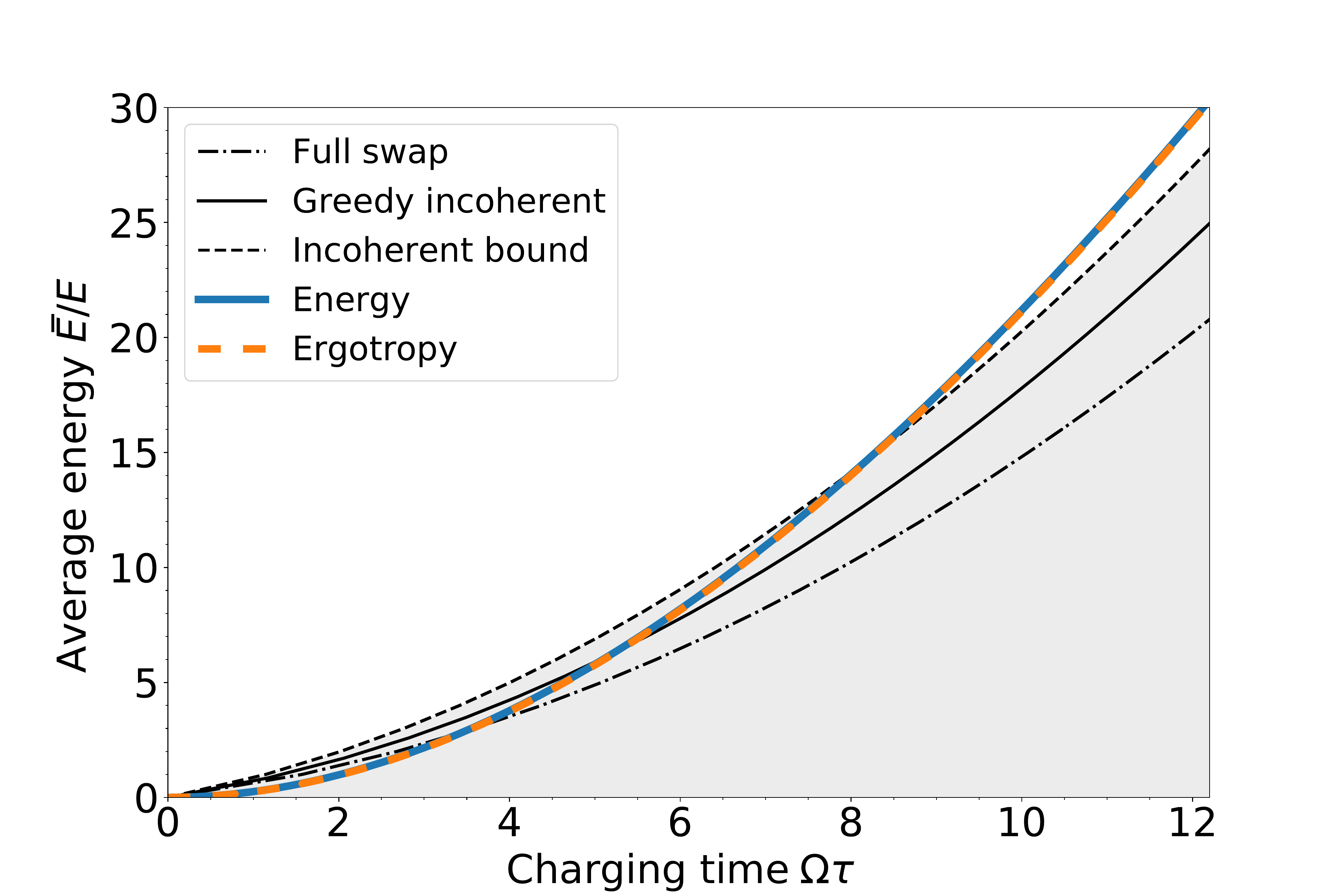}
	\caption{Collisional charging of a truncated harmonic oscillator with $250$ levels, using coherent qubits with $q=0.5$ and a swap angle $\theta = 0.01\pi$, and with a damping of $\gamma = 10^{-3}$.}
	\label{fig:coh1_loss}
\end{figure}

 Given this model of photon loss, we evaluate the cumulative charging power $\bar{\mathcal{P}}$ in \eqref{eq:DefPower} taking for $\tau$  the total time of the charging steps in which qubit and battery interact
(i.e. the preparation time of the qubit units is not accounted for in the charging power). 
The damping effect results in a correction of the incoherent charging limit \eqref{eq:Ebound_HO}, see App.~\ref{app:bound_with_dissipation} for a derivation.  Figure~\ref{fig:suptim} plots the numerically evaluated point in time at which the coherent advantage first appears (i.e.~the charging power exceeds the incoherent bound) as a function of $\theta$ for various $\gamma$-values. It quickly diverges with growing $\theta$, and as $\gamma$ reaches a critical value $\gtrsim 2 \times 10^{-3}$, above which the advantage is no longer there. For intermediate $\gamma$-values, there is an optimal $\theta$-value for achieving the earliest coherent speedup in the protocol.  We also observe that, as the best coherent charging protocols require small qubit-battery collision times $\theta$, they are also vulnerable to small losses $\gamma$. For example, choosing $\theta = 0.01\pi$, a transient coherent advantage can be observed over a a finite window of charging times for $\gamma = 10^{-3}$, as shown in Fig.~\ref{fig:coh1_loss}.

The required loss value  of  order of $10^{-3}$ can be achieved in  state-of-the-art experiments. 
In the field of cold atoms, a setup similar to Ref.~\cite{Sayrin2011} could work. In this experiment, a high-quality microwave cavity would play the role of the battery, through which a sequence of individual atoms would be sent (the charging units). The state of individual atoms can be controlled via unitary operations, allowing for incoherent and coherent charging processes by tuning the parameter $c$ as introduced in Eq.~\eqref{eq:qubitState}. Resonant interaction between the units and cavity could be implemented by choosing adequately the cavity mode. Moreover, projective energy measurements of the cavity and the charging atoms have been successfully implemented \cite{Sayrin2011} and could be used to monitor the charging process. In the experimental setup of Ref.~\cite{Sayrin2011}, a cavity with an inverse photon loss rate $1/\kappa \approx 65\,$ms was made to collide with Rydberg atoms at intervals of $t_0 = 82\,\mu$s. This corresponds to a damping factor of $\gamma \approx 1.3 \cdot 10^{-3}$ between consecutive collisions, which is sufficiently low to support an observable quantum advantage in our model. 

Superconducting platforms are also promising for demonstrating this quantum advantage. There, a microwave resonator would play the role of the battery, whereas a superconducting qubit repeatedly reinitialized in a coherent or incoherent state would play the role of the charging units. The resonant interaction between resonator and superconducting does correspond to flip-flop type interaction in the dispersive limit. As an example, we can cite a recent experiment Ref.~\cite{Linpeng2022}, where a cavity with coherence time 100 $\mu$s is coupled to a superconducting qubit. The cavity coherence time does include the cavity dissipation rate $\kappa/ 2 \pi \sim 0.5$ MHz. Initialization and interaction times are estimated to be of the order of $10^{-1} \mu$s. This leads to a rate $\gamma$ exceeding the value for which we predict a quantum advantage but technological advances in this field are promising. The coherence time of 100 $\mu$s would allow for simulating $\sim 100$ charging units.


\section{Large spin battery}
\label{section:spin}

We will now discuss coherent and incoherent charging of a large spin battery.  
Following the previous section on suitable experimental platforms, 
in this case we
bring the attention of the reader to recent hybrid experiments manipulating magnons, either within a cavity QED setup \cite{Zhang2015} or within circuit QED circuits \cite{Wolski2020}. Magnons are quanta of collective spin excitations considering ordered magnetic phases. In Ref.~\cite{Wolski2020}, they manipulate a YIG-crystal (yttrium-iron-garnet) magnetized by a uniform magnetic field. This leads to uniform magnetostatic mode (Kittel mode), which 
is coupled coherently with a superconducting transmon-type qubit. This constitutes the basic ingredients necessary for the charging protocol we propose, which we discuss in detail in what follows. 

We assume a large spin battery, with half-integer or integer quantum number $j$ and dimension $2j+1$. Its free Hamiltonian is given by the axial spin operator, $\oH_B = E \oJ_z = E \sum_{m=-j}^j m |j,m\ra$, with eigenvectors $|j,m\ra$. As before, the battery starts in the ground state $|j,-m\ra$ and gets charged by a sequence of resonant qubits with $\oH_Q = E |e\ra\la e|$, prepared in the state~(\ref{eq:qubitState}). In each collision, the qubit and the battery interact for a time $t_k$ through the energy-preserving exchange interaction Hamiltonian 
\begin{equation}
	\label{hamiltoniana_spin}
	\oH = \Omega \left( \oJ_- \otimes |e\ra\la g| + \oJ_+ \otimes |g\ra\la e| \right)
\end{equation}
where the ladder angular momentum operators are defined by
\begin{eqnarray}
\label{coupling_spin}
	\hat{J}_{\pm} \ket{j,m} &=& f_{\pm} (j,m) \ket{j,m\pm 1}, \nonumber \\
	f_{\pm} (j,m) &=& \sqrt{j(j+1)-m(m\pm 1))}.
\end{eqnarray}
We follow along the lines of Sec.~\ref{section:jaynes-cummings} and first  derive a non-tight upper bound for adaptive incoherent charging protocols, we introduce the deterministic full-swap protocol, and finally we show the coherent advantage.

\subsection{Bound on the incoherent charging power}

Starting from the ground state, $\rho_B (0)=|j, -m\ra \la j, -m|$, the state of the battery remains diagonal in the energy basis when charged by qubits without coherence. In full analogy with the rate equation \eqref{eq:MEinc_HO} for the oscillator case, and calling $\theta_k=\Omega t_k$, we can write the rates for jumping one step up and down the spin ladder as
\begin{eqnarray}
P_- (m,k) &=& q\sin^2 \left[f_-(m,k)\theta_k \right] \; , \nonumber \\
P_+ (m,k) &=& (1-q)\sin^2  \left[f_+(m,k) \theta_k \right] \; , \nonumber \\
P_0 (m,k) &=& 1-P_- (m,k)-P_+ (m,k) \; .
\end{eqnarray}
Consequently, the mean energy gain of the battery in the $k$-th incoherent charge step reads as
\begin{equation}
\Delta \bar E (k) = E\sum_{m=-j}^j \left[ P_+ (m,k)-P_- (m,k) \right] p(m,k-1)  , 
\end{equation}
which is once again optimized by taking $q=0$ such that $P_-(m,k)=0$. 
Thus, in a completely analogous way as in (\ref{eq:jensen_maxPower}), we can bound the transient charging rate of the spin battery by
	\begin{eqnarray}
	\frac{\mathcal{P}(k)}{E\Omega} &=& \frac{\Delta \bar E (k)}{E\theta_k}
	\leq R_0 \sum_{m=-j}^{j} p(m,k-1) f_+(j,m)
	 \\
	 &\leq& R_0  \sqrt{j(j+1)- \sum_{m=-j}^{j} p(m,k-1) m(m+1)} \nonumber \\ 
	 &\leq& R_0 \sqrt{j(j+1) - \tr \left\{(\oJ^2_z - \oJ_z) \rho_B(k-1) \right\} } \nonumber \\
	 &\leq& R_0 \sqrt{j(j+1) - \la J_z (k-1) \ra^2 - \la J_z (k-1) \ra } ,\nonumber
	\label{jensen_maxrate_spin}
	\end{eqnarray}
	with $\la J_z(k)\ra = \tr \{ \oJ_z \rho_B(k) \} = \bar{E}(k)/E$. Interpolating this expectation value linearly between successive steps, we arrive at the differential inequality
	\begin{equation}
	\frac{\diff \langle J_z(\Omega \tau) \rangle }{\diff \tau} \leq R_0 \Omega \sqrt{j(j+1) - \langle J_z(\Omega \tau) \rangle \left( \langle J_z(\Omega \tau) \rangle + 1\right)} .
	\label{equazione_differenziale_spin}
	\end{equation}	
	Given the initial condition $\langle J_z(0) \rangle = -j$, we can integrate \eqref{equazione_differenziale_spin} and obtain the inequality
	\begin{equation}
	\arctan\left[\frac{2\langle J_z(\Omega \tau) \rangle+1}{2f_+ \left( j,\langle J_z(\Omega \tau) \rangle \right)}\right] - \arctan\frac{1-2j}{2} \leq R_0 \Omega \tau . \label{bound_spin}
	\end{equation}	
The incoherent bound on the average charging power follows by setting both sides equal and solving for $\la J_z(\Omega \tau)\ra$, such that $\bar{\mathcal{P}} \leq E [\la J_z(\Omega \tau)\ra+j]/ \tau$.

From~(\ref{bound_spin}), we can also estimate the time $\tau^\ast$ it takes to fully charge a large-spin battery from $\la J_z(0) \ra = -j$ to $\la J_z(\tau^\ast) \ra = j$ with an incoherent protocol,
\begin{eqnarray}
\label{bound_spin_caricada0}
\tau^\ast &\geq& \frac{1}{R_0 \Omega} \left[ \arctan \frac{2j+1}{2} - \arctan\frac{1-2j}{2} \right] \nonumber \\
 && \xrightarrow{j \to \infty} \frac{\arctan(\infty)-\arctan(-\infty)}{R_0 \Omega} = \frac{\pi}{R_0 \Omega}.
\end{eqnarray}
Hence for a full charge at $j\gg 1$, the minimal time no longer depends on the number of battery levels due to the $j$-scaling of the level-dependent coupling term \eqref{coupling_spin}. Accordingly, the average power for a full incoherent charge of the battery does not exceed $\bar{\mathcal{P}}^* \leq 2 E j \Omega R_0/\pi$.

\subsection{Classical strategies}

As in the oscillator case, we can consider a full-swap strategy in which we adapt the coupling times according to $\vartheta_k = \pi/2f_+(j,k)$, such that the battery is always in a pure charge state $|j,k-j\ra$. The energy and ergotropy are equal and  increase in steps of $E$. In order to completely charge the battery, $2j$ such swaps must be performed, which requires the total time 
\begin{eqnarray}
\tau^* = \frac{1}{\Omega} \sum_{k = -j}^{j-1} \frac{\pi}{2f_+(j,k)} \xrightarrow{j \to \infty} \frac{\pi^2}{2\Omega} .
\end{eqnarray}
At $j\gg 1$, the overall charging power $\bar{\mathcal{P}} = 2Ej/\tau^*$ is then $2/R_0\pi \simeq 0.88$ times lower than the incoherent upper bound deduced from (\ref{bound_spin_caricada0}).

A slightly better performance is obtained by the \emph{greedy} incoherent protocol (shown in Fig.~\ref{fig:coh1_spin}), which  selects as the swap angle $\theta_k$ in each step the local maximum of the cumulative average charging power. 

\subsection{Coherent charging}

\begin{figure}
	\centering
	\includegraphics[width=\columnwidth]{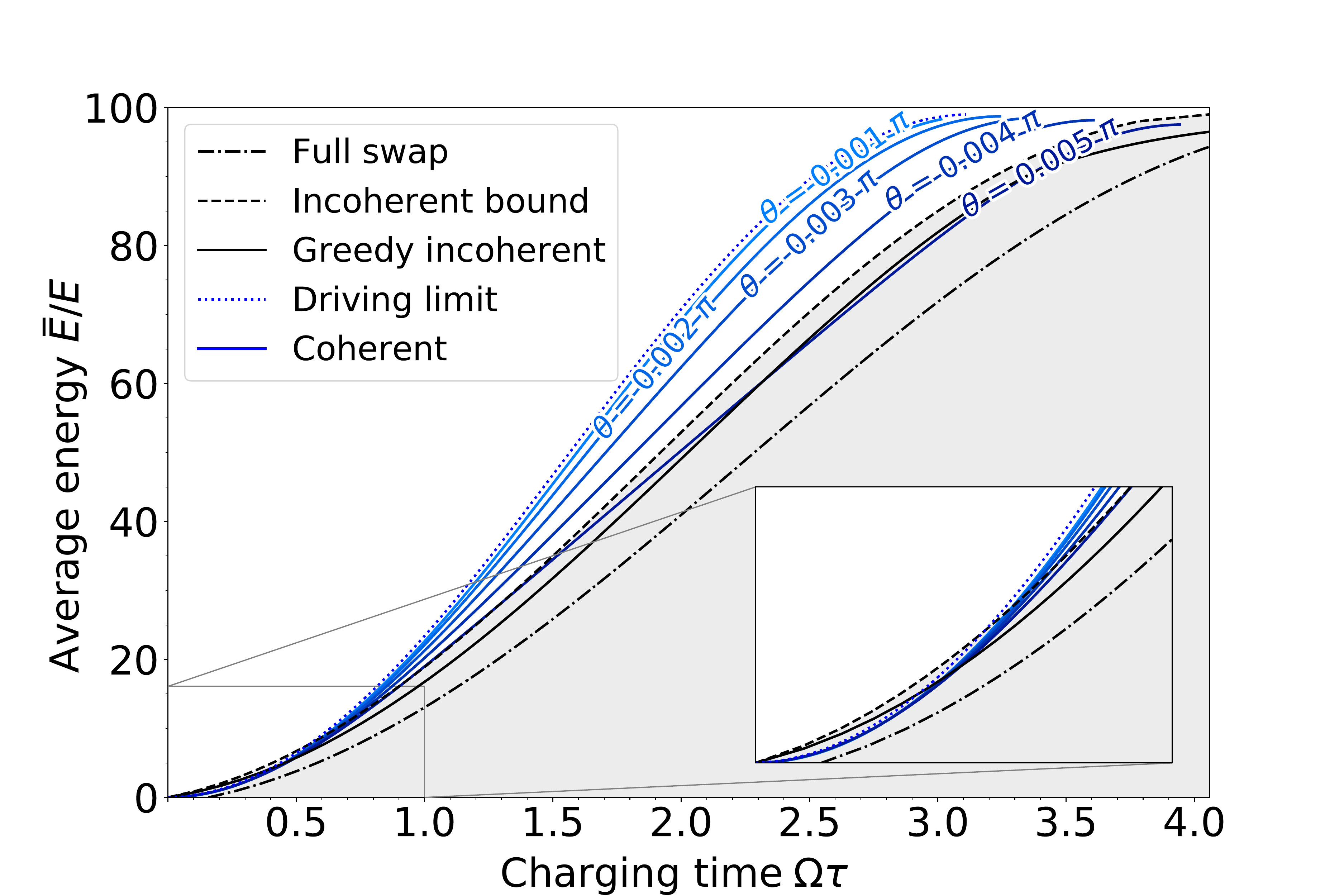}
	\caption{Mean battery energy $\bar{E}$ as a function of charging time $\Omega \tau = k\theta$, for a spin-$99/2$ battery with $100$ levels charged by a sequence of qubits. Blue solid lines of darker shades correspond to coherent qubits at $q=0.5$ using increasing swap angles $\theta$, and the blue dotted line corresponds to the driving limit $\theta \to 0$. The black dashed line marks the upper bound for adaptive incoherent charging strategies with excited qubits ($q=0$), which is almost reached by a greedy protocol (black solid line) optimizing the cumulatively averaged power at every charge step. The dash-dotted line shows the less performant deterministic full-swap protocol. Inset: Zoom on the time interval where the coherent protocol for the charging process becomes better than the incoherent bound.}
	\label{fig:coh1_spin}
\end{figure}

The minimum full-charge time \eqref{bound_spin_caricada0} and the associated power bound hold for all incoherent protocols, but they can be overcome with coherent qubits. As for the oscillator battery, the optimal performance is obtained with the choice $q=1/2$ and $c=1$, regardless of the phase angle $\alpha$. 
As an example, we compare coherent and incoherent charging protocols for a spin battery of dimension $d=2j+1=100$ in Fig.~\ref{fig:coh1_spin}. We plot the average energy relative to the ground state, $\bar{E}(K)+Ej$, as a function of the charging time $\Omega \tau = \sum_{k=1}^K \theta_k$. The blue solid curves from light to dark correspond to non-adaptive coherent charging protocols ($q=1/2$, $c=1$, $\alpha=0$) at growing values of the fixed swap angle $\theta_k=\theta$; they beat the incoherent bound (black dashed) when $\theta \lesssim 0.004\pi$. The optimized greedy incoherent protocol (black solid) gets close to the bound and clearly outperforms the full-swap protocol (dash-dotted).
	
The blue dotted line represents  the asymptotic coherent driving limit $\theta \to 0$ in which the battery state remains pure and evolves according to
\begin{equation}
    \rho_B (K) \xrightarrow[K\to\infty]{\theta\to 0} \oD(\tau) |j,-j\ra \la j,-j| \oD^\dagger(\tau).
\end{equation}
Here, the unitary describes a coherent spin rotation by the angle $\Omega \tau$ about the $x$-direction on the generalized Bloch sphere,
\begin{eqnarray}
\oD(\tau) = \exp \left[ \frac{\Omega \tau}{2}\left( \hat{J}_+ + \hat{J}_- \right) \right].
\end{eqnarray}
A complete charge of the battery corresponds to a $\pi$-rotation, which requires the time $\tau^\star = \pi / \Omega$ and is therefore $R^{-1}_0 \simeq 1.38$ times faster than the incoherent bound~\eqref{bound_spin_caricada0}, corresponding to a $38\%$ quantum advantage.

\section{Conclusions}
\label{section:conclusions}

Despite immense  progress on our understanding of thermodynamics processes in the quantum regime~\cite{Goold2016,Vinjanampathy2016,Mitchison2019,myers2022quantum}, identifying  quantum advantages in thermodynamics  remains a challenging task. Quantum batteries offer a natural framework to explore this question, and previous efforts focused on  quantum advantages of a collective nature based on entanglement~\cite{Campaioli2018,Bhattacharjee2021,Alicki2013,Hovhannisyan2013,Binder2015,Campaioli2017,JuliaFarre2020,Gyhm2022}. Building upon our previous work~\cite{Seah2021}, here we have shown that such quantum advantages can also appear at the level of a single battery: given a collisional model of charging, we have proven that quantum coherent charging processes can outperform arbitrary classical charging strategies. More precisely, we have considered a set of qubit units that interact sequentially with  a quantum battery, given by either a quantum harmonic oscillator (thus realizing  a micromaser) or a large-spin system. A quantum advantage arises in both set-ups,  
with 
90\% and 38\% higher charging power through quantum resources for the micromaser and the spin batteries, respectively. We have characterised the robustness of the quantum advantage to decoherence and photon loss, and discussed its feasibility in both cavity QED~\cite{Sayrin2011} and hybrid superconducting ~\cite{Linpeng2022,Wolski2020} set-ups.  An experimental demonstration of these results is an exciting possibility that we hope can be achieved in the near future. \\

\textit{Note added.--} Upon finalising this work, Ref.~\cite{Shaghaghi2022} appeared, which is also concerned by charging an oscillator battery with a sequence of coherent qubits in a micromaser set-up. Complementary to our results, the authors focus on the regime of ultrastrong cavity-qubit coupling and include counter-rotating interaction terms to their model. 



\section{Acknowledgements}

We thank Kater Murch, Vittorio Giovannetti and Stella Seah for discussions.  We acknowledge funding from the Swiss National Science Foundation (NCCR SwissMAP). G. H. and M. P.-L.  acknowledge the Swiss NSF for financial support through the starting grant PRIMA PR00P2$\_$179748 and the Ambizione grant PZ00P2-186067.

\bibliography{references}

\newpage

\appendix

\begin{widetext}

\section{Master equation model for sequential battery charging}
\label{app:dissipatori}

Here we  present a generic master equation model for the sequential charging of a ladder battery by identical qubits as described in the Framework, Sec.~\ref{section:framework}.

\subsection{State change per charging step}

Each charging step is represented by a unitary operation $\oU$ acting on the combined Hilbert space of battery and qubit, which results in an effective quantum channel $\rho_B \mapsto \Phi \rho_B$, repeatedly applied to the reduced battery state. After $k$ steps, we have $\rho_B(k) = \Phi^k \rho_B(0) = (1+\cL)^k \rho_B(0)$. Given a Kraus representation of the channel, $\Phi \rho_B = \sum_i \oM_i \rho_B\oM_i^\da$ with $\sum_i \oM_i^\da \oM_i = \id$, we can express the incremental change of state $\cL$ as a sum of Lindblad dissipators,
\begin{equation}
    \cL \rho_B := \Phi\rho_B - \rho_B = \sum_i \cD[\oM_i] \rho_B,
\end{equation}
using the abbreviation $\cD[\oM]\rho = \oM\rho\oM^\da - \{\oM^\da\oM,\rho\}/2$.



Here, one obtains a valid set of Kraus operators from the eigenbasis of the qubit state, $\rho_Q = \sum_{n=0}^1 q_n |\psi_n\ra\la \psi_n|$, and an auxiliary orthonormal basis $\{|\chi_0\ra, |\chi_1\ra \}$,
\begin{equation}
    \Phi \rho_B = \tr \left\{ \oU \rho_B \otimes \rho_Q \oU^\da \right\} = 
    \sum_{m,n=0}^1 \oM_{m,n} \rho_B \oM_{m,n}^\da \; , \qquad \oM_{m,n} = \sqrt{q_n} \la \chi_m|\oU| \psi_n \ra.
\end{equation}
One can easily check that $\sum_{m,n} \oM_{m,n}^\da \oM_{m,n} = \id$, given $q_0+q_1 = 1$. This determines the master equations for incoherent and coherent charging, as expressed by the Lindblad generators $\cL_{\rm{inc}}$ and $\cL_{\rm{coh}}$. 

For the incoherent case, the qubits are prepared in a mixed state $\rho_Q^{\rm{(inc)}}$ of a given ground-state probability $q$, i.e., $|\psi_0\ra = |g\ra$, $|\psi_1\ra = |e\ra$, and $q_0 = q$. Using the same auxiliary basis, we are left with
\begin{equation}
    \cL_{\rm{inc}} \rho_B = q \left( \cD[\la g|\oU|g\ra] + \cD[\la e|\oU|g\ra] \right)\rho_B + (1-q) \left( \cD[\la g|\oU|e\ra] + \cD[\la e|\oU|e\ra] \right)\rho_B . 
\end{equation}
The amount of \textit{useful} energy each qubit can at most transfer to the battery is specified by the ergotropy $\cE_Q = \max \{0,1-2q\}E \geq 0$. It is lower than the mean energy $(1-q)E$ and finite only for population-inverted states ($q<1/2$).

For the coherent case of the same mean energy, the qubits are in a pure superposition state $\rho_Q^{\rm{(coh)}} = |\psi_0\ra\la \psi_0|$, with $|\psi_0\ra = \sqrt{q} |g\ra + \sqrt{1-q} e^{i\alpha} |e\ra$ and $q_0 = 1$. The ergotropy equals the mean energy, $\cE_Q = (1-q)E$, and is non-zero also for $q>1/2$. In order to obtain the generator, we could construct an orthogonal complement to the qubit state, $|\psi_1\ra = \sqrt{1-q} |g\ra - \sqrt{q} e^{i\alpha} |e\ra$, so that $\cL_{\rm{coh}} \rho_B = \cD[\la 0|\oU|0\ra]\rho_B + \cD[\la 1|\oU|0\ra]\rho_B$. However, it turns out to be more pertinent to use the energy states as the auxiliary basis and write
\begin{equation}
    \cL_{\rm{coh}} \rho_B = \cD[\la g|\oU|0\ra]\rho_B + \cD[\la e|\oU|0\ra]\rho_B.
\end{equation}
This form ensures that, for the energy-preserving exchange interactions discussed next, the first dissipator can only contain excitations of the battery state, while the second dissipator can contain de-excitations only.
Depending on the interaction model $\oU$, one might be able to single out an explicit coherent driving term in the form of a Hamiltonian contribution to the master equation, with help of the identity
\begin{equation}\label{eq:LindbladExpandRule}
    \cD[\oB + \alpha \id] \rho = \cD[\oB]\rho + \left[ \frac{\alpha^{*}\oB - \alpha \oB^\da}{2},\rho \right].
\end{equation}

Partially coherent qubit states of a given ground-state population $q$ can always be written as a convex combination of the corresponding mixed and pure states, $\rho_Q = c\rho_Q^{\rm{(coh)}} + (1-c)c\rho_Q^{\rm{(inc)}}$ with $c \in [0,1]$. This implies the same convex combination for the charging generator, $\cL = c\cL_{\rm{coh}} + (1-c)\cL_{\rm{inc}}$. Ergotropy is then provided in the form of both coherence ($c>0$) and population inversion ($q<1/2$), since $\cE_Q = E [1-2q + \sqrt{(1-2q)^2 + 4c^2q(1-q)}]/2$.

\subsection{Energy-preserving resonant exchange}

In our charging model, we demand that the battery charge be sourced exclusively by the charging qubits and that no additional work be consumed in the process. Hence, the interaction must be energy-preserving, $[\oU,\oH_B + \oH_Q] = 0$, and the associated quantum channel $\Phi$ a so-called thermal operation \cite{Brandao2013,Brandao2015}. 

Specifically, we restrict our view to the resonant exchange of single excitations, as described by partial swap operations between the qubit and neighbouring battery levels, $\oU = \exp(-i\oH \theta)$ generated by a Hermitian coupling Hamiltonian of the form
\begin{equation}\label{eq:Hint2}
    \oH = \oA \otimes |e\ra\la g| + \oA^\da \otimes |g\ra\la e| , \quad \la m|\oA|n\ra \propto \delta_{m,n-1}.
\end{equation}
Here, $\oA$ represents a model-dependent ladder operator moving downward on the battery, while the parameter~$\theta$ characterizes the level-dependent swap angle (including the coupling frequency $\Omega$). For the studied cases of an oscillator and a spin battery in Sec.~\ref{section:jaynes-cummings} and \ref{section:spin}, we used $\oA = \oa$ and $\oA = \oJ_-$, respectively.
Note that the most general energy-preserving $\oH$ could also contain phase rotation terms of the form $\sum_n |n\ra\la n| \otimes (v_{n,e} |e\ra\la e| + v_{n,g} |g\ra\la g|)$. Such terms would complicate the charging dynamics and are not considered here.

The unitary charging operation can be expanded as 
\begin{eqnarray}
    \oU &=& e^{-i\oH\theta} 
    = \cos (|\oH|\theta) - i\theta \sinc (|\oH|\theta) \oH,
\end{eqnarray}
introducing the absolute value operator 
\begin{equation}
    |\oH| := \sqrt{\oH^2} = \sqrt{\oA\oA^\da \otimes |e\ra\la e| + \oA^\da \oA \otimes |g\ra\la g|} \; .
\end{equation}
It satisfies $[\oH, |\oH| ] = 0$ and it is also diagonal in the product basis of battery and qubit energy states, because $\oA^\da\oA$ and $\oA\oA^\da$ are diagonal in the basis of battery levels $|n\ra$ by construction. However, $|\oH|$ will not be full rank for (semi-)finite batteries with empty and fully charged states at $n=0$ and $n=N$ (or $n\to \infty)$, since $|0,g\ra$ and $|N,e\ra$ are not affected by $\oH$. 

From the matrix elements of the unitary, we can identify Lindblad operators $\oL_{-}$ and $\oL_{+}=\oL_{-}^\da$ associated to jumps down and up the battery ladder with level-dependent jump probabilities, as well as Lindblad operators $\oL_{e,g} = \oL_{e,g}^\da$ associated to dephasing between energy levels,
\begin{eqnarray}
\oL_g = \la g |\oU |g\ra &=& \cos (\sqrt{\oA^\da \oA} \theta), \nonumber \\
\oL_e = \la e |\oU |e\ra &=& \cos (\sqrt{\oA \oA^\da} \theta), \nonumber \\
\oL_- = i \la e |\oU |g\ra &=& \theta \sinc (\sqrt{\oA \oA^\da} \theta) \oA = \theta \oA \sinc (\sqrt{\oA^\da \oA} \theta), \nonumber \\
\oL_+ = i \la g |\oU |e\ra &=& \oL_-^\da. 
\end{eqnarray}
In the third line, we make use of the fact that $(\oA\oA^\da)^k\oA = \oA (\oA^\da\oA)^k$ for $k \in \mathbb{N}$ due to associativity. The generators for incoherent and coherent charging become
\begin{eqnarray}
    \cL_{\rm{inc}} \rho_B &=& q \left( \cD[\oL_g] + \cD[\oL_-] \right)\rho_B  + (1-q) \left( \cD[\oL_e] + \cD[\oL_+] \right)\rho_B , \label{eq:MEinc} \\
    \cL_{\rm{coh}} \rho_B &=& \cD \left[ \sqrt{q}\oL_- + i\sqrt{1-q}e^{i\alpha}\oL_e \right]\rho_B + \cD \left[ \sqrt{1-q}\oL_+ + i\sqrt{q}e^{-i\alpha}\oL_g \right]\rho_B. \label{eq:MEcoh}
\end{eqnarray}
For incoherent charging, the dephasing terms $\oL_{e,g}$ can be ignored if there are no energy coherences in the battery state to begin with. The coherent process, on the other hand, will build up such coherences.

For small $\theta$, we can expand the generators to second order in $\theta$. By virtue of the identity \eqref{eq:LindbladExpandRule}, we arrive at 
\begin{eqnarray}
    \cL_{\rm{inc}} \rho_B &\approx& q \theta^2 \cD[\oA] \rho_B + (1-q) \theta^2 \cD[\oA^\da] \rho_B, \label{eq:MEinc_small} \\
    \cL_{\rm{coh}} \rho_B &\approx& \cL_{\rm{inc}} \rho_B -i \sqrt{q(1-q)}\theta \left[ e^{-i\alpha} \oA + e^{i\alpha}\oA^\da,\rho_B \right]. \label{eq:MEcoh_small}
\end{eqnarray}
That is to say, in the limit of short or weak interactions, the incoherent charging process is always of second order in $\theta$, whereas the coherent process adds an effective driving Hamiltonian $\oV = \sqrt{q(1-q)}\theta e^{-i\alpha} \oA + h.c$. The driving term is of first order in $\theta$, and so it dominates whenever $q\neq 0,1$. It preserves the purity of the battery state and is responsible for the coherent advantages we have discussed for oscillator and spin batteries.

In Appendix~\ref{app:ladder}, we derive explicit Lindblad operators for a uniform ladder battery and recover the random-walk charging process studied in~\cite{Seah2021}. There, we can make use of \eqref{eq:LindbladExpandRule} to single out a Hamiltonian driving term for the coherent case.

\section{Random-walk battery}
\label{app:ladder}

Let us use our generic results and check consistency with the random-walk battery we have studied before in \cite{Seah2021}. The uniform ladder operator is $\oA = \sum_{n=1}^N |n-1\ra\la n|$, and we can quickly convince ourselves that 
\begin{equation}\label{eq:RWladder_AdA}
    \oA^\da \oA = \id - |0\ra\la 0| = \sqrt{\oA^\da \oA}, \quad \oA \oA^\da = \id - |N\ra\la N| = \sqrt{\oA \oA^\da}.
\end{equation}
In particular, these projectors imply that $f(\oA\oA^\da)\oA = f(\id) \oA$ and $f(\oA^\da\oA)\oA^\da = f(\id) \oA^\da$. Hence we get for the incoherent model
\begin{eqnarray}
\cL_{\rm{inc}} \rho_B &=& q \sin^2 \theta \cD \left[ \oA \right] \rho_B  + (1-q)\sin^2\theta \cD \left[ \oA^\da \right]  \rho_B \nonumber \\
&& + \; q \cD \left[ |0\ra\la0| + (\id - |0\ra\la0|)\cos\theta \right] \rho_B + (1-q) \cD \left[ |N\ra\la N| + (\id - |N\ra\la N|)\cos\theta \right] \rho_B \nonumber \\
&=& \sin^2 \theta \left( q\cD [ \oA ]  + (1-q)\cD [ \oA^\da ]\right)  \rho_B  + (1-\cos\theta)^2 \left( q\cD [ |0\ra\la 0| ] + (1-q)\cD [ |N\ra\la N| ] \right)\rho_B , \nonumber \\
\end{eqnarray}
where for the last line we made use of \eqref{eq:LindbladExpandRule}. The result is exactly what we got in \cite{Seah2021}. In particular, the dephasing only affects the two boundary levels directly. 

For the coherent model, we similarly get
\begin{eqnarray}
\cL_{\rm{coh}} \rho_B &=& \cD \left[ \sqrt{q} \sin\theta \oA  +  i\sqrt{1-q}e^{i\alpha} (\cos\theta \id + (1-\cos\theta)|0\ra\la 0| )  \right] \rho_B \nonumber \\
&& + \; \cD \left[ \sqrt{1-q} \sin \theta \oA^\da + ie^{-i\alpha}\sqrt{q} (\cos\theta \id + (1-\cos\theta)|N\ra\la N| ) \right] \rho_B \nonumber \\
&=& -i \sqrt{q(1-q)}\sin\theta\cos\theta \left[ e^{-i\alpha}\oA + e^{i\alpha}\oA^\da,\rho_B \right] \nonumber \\
&& + \; \cD \left[ \sqrt{q} \sin\theta \oA +  i\sqrt{1-q}e^{i\alpha} (1-\cos\theta)|0\ra\la 0|  \right] \rho_B  + \cD \left[ \sqrt{1-q} \sin \theta \oA^\da + ie^{-i\alpha}\sqrt{q} (1-\cos\theta)|N\ra\la N| \right] \rho_B . \nonumber \\
\end{eqnarray}
Using again \eqref{eq:LindbladExpandRule} for the identity term inside the dissipators has lead us to our old result. However, we see that the appearance of an explicit driving Hamiltonian is due to the simple projectors \eqref{eq:RWladder_AdA} and cannot be expected for other models.

\section{Incoherent charging bound with dissipation}
\label{app:bound_with_dissipation}

Here we derive an upper bound for the incoherent charging of an oscillator battery in the presence of damping. Our model in  Sec.~\ref{sec:experiment} assumes that the charging events are short and separated by a waiting time $\tau_0$ during which the oscillator is subject to a (quantum-limited) attenuation channel \eqref{eq:damping} with a damping constant $\gamma$.

We can restrict our view to excited qubits ($q=0$), so that, in the charge step $(\rm{ch})$, the battery either gains one charge unit $E$ with probability $P_+ (n) = \sin^2 (\sqrt{n+1}\theta)$ or remains at its current level $n$ with $P_0(n) = 1-P_+(n)$. During the damping period that follows, the battery can only lose charge. Starting from the $n$-th level, the energy loss due to damping is $\Delta E^{(\rm{da})} (n) = E \tr [\oa^\da \oa \Phi_0 (|n\ra\la n|)] - E n = -\gamma E n$. Hence, similar to \eqref{eq:DeltaE_inc_HO}, the combined mean energy change over the $k$-th charging step with swap angle $\theta_k$ and the subsequent damping can be written as
\begin{equation}
    \Delta \bar{E}(k) = E \sum_{n=0}^\infty p(n,k-1) \left[ P_+(n,k) [1-\gamma (n+1)] - P_0(n,k) \gamma n \right] = E \sum_{n=0}^\infty p(n,k-1) \left[ (1-\gamma) \sin^2 (\sqrt{n+1}\theta_k) - \gamma n \right].
\end{equation}
In full analogy to Sec.~\ref{section:jaynes-cummings}, we can now construct an upper bound on the transient charging power $\mathcal{P}(k) = \Omega \Delta \bar{E}(k)/\theta_k $ by maximizing and upper-bounding each $n$-summand individually,
\begin{equation}
	\max_{\theta \in [0, \frac{\pi}{2}]} \frac{(1-\gamma) \sin^2 (\sqrt{n+1}\theta) - \gamma n}{\theta}     \leq (1-\gamma) \max_{\theta \in [0, \frac{\pi}{2}]} \frac{ \sin^2(\sqrt{n+1}\theta)}{\theta} -  \gamma \min_{\theta \in [0, \frac{\pi}{2}]} \frac{n}{\theta} = (1-\gamma)\sqrt{n+1}R_0 - \frac{2 \gamma n}{\pi} .
	\label{eq:maxrate_lossy}
\end{equation}
Note that, for the sake of clarity, we are only counting the charging time and omitting the damping time $t_0$ here. Otherwise, the overall charging time would be offset by $Kt_0$.
Then equation \eqref{eq:jensen_maxPower} has to be modified to
	\begin{eqnarray}
	\frac{\Delta \bar E (k) }{E\theta_k} \leq  R_0 (1-\gamma) \sqrt{\frac{\bar E(k-1)}{E} + 1} - \frac{2}{\pi} \gamma \bar E(k-1) ,
	\label{jensen_maxrate_lossy}
	\end{eqnarray}	
	and therefore (assuming a growing $\bar E$),
	\begin{eqnarray}
	\frac{\diff \bar E (\Omega \tau) }{\diff \tau} \leq  \Omega E R_0 (1-\gamma) \sqrt{\frac{\bar E(\Omega \tau)}{E} + 1} - \frac{2\Omega}{\pi} \gamma \langle E(\Omega \tau) \rangle \; .
	\label{equazione_differenziale_lossy}
	\end{eqnarray}	
	The solution of \eqref{equazione_differenziale_lossy} gives a bound for the incoherent charging in presence of dissipation; it does not differ appreciably from \eqref{eq:Ebound_HO} in the considered weak-damping regime $\gamma \ll 1$.






\end{widetext}

\end{document}